\documentclass[twocolumn]{openjournal} 

\usepackage[T1]{fontenc}
\usepackage{ae,aecompl}

\usepackage{graphicx}	% Including figures
\usepackage{amsmath}	% Advanced maths commands
\usepackage{amssymb}	% Extra maths symbols
\usepackage{natbib}

\usepackage{hyperref}
\hypersetup{
	colorlinks=true,
	urlcolor=blue,    % color of external links
	linkcolor=black,  % color of toc, list of figs etc.
	citecolor=blue,   % color of links to bibliography
}

\defcitealias{Pittordis_2018}{PS18}
\defcitealias{Pittordis_2019}{PS19}
\defcitealias{Pittordis_2022}{PS22} 
\defcitealias{Badry_2021}{ERH}
\defcitealias{Banik_2018_Centauri}{BZ18}

\newcommand{\kms} {{\rm \, km \, s^{-1} }} 

\newcommand{\au} {\, {\rm AU}}   
\newcommand{\yr}{\,{\rm yr}} 
\newcommand{\kau} {\, {\rm kAU}} 
\newcommand{\msun} {\,M_\odot} 
\newcommand{\masyr} {\, {\rm mas}\, {\rm yr}^{-1} } 
\newcommand{\muasy} {\, \mu{\rm as} \, {\rm yr}^{-1} } 
\newcommand{\muas}{\, \mu{\rm as} } 
\newcommand{\pc} {\, {\rm pc}} 
 
\newcommand{\msecsq}{\, {\rm m \, s^{-2}}}
\newcommand{\logten}{\log_{10}}

\renewcommand{\bv}{\mathbf{v}} 
 
\newcommand{\vtilde}{\tilde{v}}

\newcommand{\newa}{  }  

\begin{document}
	
	\title{Wide Binaries as a modified gravity test: prospects for detecting triple-system contamination}

	\author{D. Manchanda} 
	\email[E-mail: ]{d.manchanda@qmul.ac.uk}
	\author{W. Sutherland} 
	\email[E-mail: ]{w.j.sutherland@qmul.ac.uk}
		\author{C. Pittordis} 
	\email[E-mail: ]{c.pittordis@qmul.ac.uk}
	\affiliation{The School of Physical and Chemical Sciences,  
		Queen Mary University of London, Mile End Road, London E1 4NS, UK.}	
	
	% Abstract of the paper
	\begin{abstract}
		Several recent studies have shown that velocity differences of very wide binary stars, measured to high precision with GAIA, can potentially provide an interesting test for modified-gravity theories which attempt to emulate dark matter; in essence, 
		MOND-like theories (with external field effect included) predict that wide binaries (wider than $\sim 7\kau$)  
		should orbit $\sim 15\%$ faster than Newtonian 
		for similar orbit parameters; such a shift is readily detectable in principle  
		in the sample of 9,000 candidate systems selected from GAIA EDR3 by Pittordis and Sutherland (2022; PS22).  
		However, the main obstacle at present is the observed ``fat tail" of candidate wide-binary systems 
		with velocity differences at $\sim 1.5 - 6 \times$ 
		circular velocity; this tail population cannot be bound pure binary systems, 
		but {\newa a possible explanation of the tail is} triple or quadruple systems
		with unresolved or undetected additional star(s).  While this tail can be modelled and statistically subtracted, 
		obtaining a well-constrained and accurate  model for the triple population is crucial to obtain a
		robust test for modified gravity.   
		In this paper we explore prospects for observationally constraining the triple population: we 
		simulate a population of hierarchical triple systems ``observed" as in PS22 at random epochs and viewing angles; 
		for each simulated system we then evaluate various possible 
		methods for detecting the third star, including GAIA astrometry, radial velocity drift, and several imaging methods from 
		direct Rubin images, speckle imaging and coronagraphic imaging.  Results are generally encouraging, in that typically 
		90 percent of the triple systems in the key regions of parameter space are detectable; 
		there is a moderate ``dead zone" of cool brown-dwarf companions at $\sim 25-100 \au$ separation 
		which are not detectable with any of our baseline methods.   A large but feasible 
		observing campaign can greatly clarify the effect of the triple/quadruple population and make the gravity test decisive.    
	\end{abstract}


	\maketitle
	
	\section{Introduction}
\label{sec:intro} 

A number of recent studies have shown that velocity differences of wide stellar binaries offer an interesting
test for modified-gravity theories similar to MoND, which attempt to eliminate the need for dark matter
(see e.g. 	\citet{Hernandez_2012},
\citet{Hernandez_2012_2}
\citet{Hernandez_2014},
\citet{Matvienko_2015}, 
\citet{Scarpa_2017} 
and \citet{Hernandez_2019}). 
Such theories 
require a substantial modification of  standard GR below a characteristic acceleration threshold 
$a_0 \sim 1.2 \times 10^{-10} \msecsq$
 (see review by \citet{Famaey_2012}, also {\newa \citet{Barrientos_2018b} and \citet{Gueorguiev_2022} } .    
A key advantage of wide binaries is that at separations $\ga 7 \kau$, the relative accelerations are 
below this threshold, so MoND-like theories predict significant deviations from GR; while wide binaries 
should contain neglible dark matter, so DM theories predict
no change from GR/Newtonian gravity. Thus in principle the predictions of DM vs modified gravity in wide binaries are
unambiguously different, unlike the case for galaxy-scale systems where the DM distribution is uncertain.   

Wide binaries in general have been studied since the 1980s (\citep{Weinberg_1987, Close_1990}), but until recently the
precision of ground-based proper motion measurements was a serious limiting factor: wide binaries could
be reliably selected based on similarity of proper motions, 
see e,g, 
\citet{Yoo_2003},
\citet{Lepine_2007},
\citet{Kouwenhoven_2010},
\citet{Jiang_2010},
\citet{Dhital_2013}, 
\citet{Coronado_2018}. 
However, the typical proper motion precision $\sim 1 \masyr$ from ground-based
or Hipparcos measurements was usually not good enough
to actually measure the internal velocity differences, except for a limited number of nearby systems.  

The launch of the GAIA spacecraft \citep{Gaia_2016} in 2014 offers a spectacular improvement in precision; the proper motion 
precision of order $30 \muasy$ corresponds to transverse velocity precision $0.0284 \kms$ at distance 200 parsecs, 
around one order of magnitude below wide-binary orbital velocities, so velocity differences can be measured
to good precision over a substantial volume; and this will steadily improve with future GAIA data extending 
eventually to a 10-year baseline. 
Recent studies of WBs from GAIA include e.g. \citet{Badry_2021} and \citet{Hernandez_2022}. 

In earlier papers in this series, \citet{Pittordis_2018} (hereafter \citetalias{Pittordis_2018}) compared 
simulated WB orbits in MoND versus GR, to investigate prospects for the test in advance of GAIA DR2.  
This was applied to a sample of candidate WBs selected from 
GAIA DR2 data by \citet{Pittordis_2019} (hereafter \citetalias{Pittordis_2019}),
and an expanded sample from GAIA EDR3 by \citet{Pittordis_2022} (hereafter \citetalias{Pittordis_2022}).   
To summarise results, simulations show that (with MoND external field effect included), wide binaries
at $\ga 10 \kau$ show orbital velocities typically 15 to 20 percent faster in MOND than GR, at equal
separations and masses.  This leads to a substantially larger fraction of ``faster" binaries with observed 
velocity differences between 1.0 to 1.5 times the Newtonian circular-orbit value. In Newtonian gravity, 
changing the eccentricity distribution changes the shape of the distribution mainly at lower velocities, 
but has little effect on the distribution at the high end from 1.0 to 1.5 times circular velocity.  
Therefore, the predicted shift from MOND is distinctly different 
from changing the eccentricity distribution within Newtonian gravity; so
given a large and pure sample of several thousand WBs
with precise 2D velocity difference measurements, 
we could decisively distinguish between GR and MOND predictions. 

The main limitation at present is that \citetalias{Pittordis_2019} and \citetalias{Pittordis_2022} showed the presence of a ``fat tail" of candidate binaries
with velocity differences $\sim 1.5$ to $6 \times$ 
the circular-orbit velocity; {\newa at ratios $\ga 2.5$, these systems are too fast to be explained by modified gravity 
 since a modification so large would overshoot galaxy rotation curve data}. A possible explanation 
\citep{Clarke_2020} is
higher-order multiples e.g. triples where either one star in the observed ``binary" is itself an unresolved closer binary, 
or the third star is at resolvable separation but is too faint to be detected by GAIA; the third star on a closer orbit thus  substantially boosts the velocity difference of the two observed stars in the wide ``binary".   

In \citetalias{Pittordis_2022} we made a simplified model of this triple population, then fitted the full distribution
of velocity differences for WB candidates using a mix of binary, triple and flyby populations. These fits found
that GR is significantly preferred over MOND {\bf if} the rather crude PS22 triple model {\newa (with equal populations
 of triples and pure 2-star binaries)}  is correct, 
but we do not know this at present.  
Allowing much more freedom in the triple modelling is computationally 
expensive due to many degrees of freedom, 
and is likely to lead to significant degeneracy between gravity modifications and varying the triple population.  
Therefore, observationally constraining the triple population, or eliminating most of it by additional observations, is the next key step  to make the WB gravity test more secure. 

In this paper we explore prospects for observationally constraining the triple population: we
generate simulated triple systems ``observed" at random epochs, inclinations and viewing angles, 
and then test whether the presence of the third star is detectable by any of various methods 
including direct, speckle or coronagraphic imaging; radial velocity drift; or astrometric non-linear motion
in the future GAIA data; we see below that prospects are good, in that 80 to 95\% of triple 
systems in the PS22 sample should be potentially detectable as such by at least one of the methods.  

The plan of the paper is as follows: in Section~\ref{sec:trip-sims} we 
outline the parameters (semi-major axis distributions, eccentricities, masses etc) 
used in the simulated triple systems.  
In Section~\ref{sec:tripdet} we describe
the various methods and thresholds adopted for defining a simulated detection of the third star. 
In Section~\ref{sec:results} we show various results and plots, indicating which methods are successful at detecting
a third object in various regions of observable parameter space, in particular as a function of projected velocity (relative to circular-orbit value) and projected separation. 
We summarise our conclusions in Sec.~\ref{sec:conc}.

%% move these to intro - done 	

%%%  simulations %%%

\section{Triple simulations} 
\label{sec:trip-sims} 

In general the detectability of a third star in a candidate ``wide binary" depends on many parameters including mass ratio, orbit size, eccentricity, inclination and phase, and the system distance; this is in principle calculable analytically, but the calculations are complex so this is best handled by a Monte-Carlo type simulation as here; we set up a large number of simulated triple systems and take ``snapshots" of these at random times and viewing angles, 
then evaluate detectability of the third star for each simulated  snapshot.  

In this section we describe the parameters and distribution functions used to set up simulated triple systems.
For simplicity, we model each triple system as two independent Kepler orbits, ``Outer" and ``Inner"; 
we label the stars so that stars 2 and 3 are the inner binary, with $M_2 > M_3$, 
while the ``outer" orbit consists of star 1 and
the barycenter of stars 2 and 3; subscripts ``inn" and ``out" below denote these.   Thus, the ``wide binary" 
observed in PS22 comprises star 1 plus either star 2, or the unresolved blend of stars 2+3. This wide system is always
well resolved, since the PS22 subsample used in fitting has angular separations $\ge 5 \kau / 300 \pc = 16.6\, $arcsec; 
the inner pair may be resolved or unresolved, as below.  

We choose semimajor axes so the distribution of $a_{out}$ is uniform in $\logten a_{out}$ 
from $0.1 \kau \le a_{out} \le 100 \kau$; this is chosen significantly wider than the range of 
interest $ 5 \le r_p \le 20 \kau$ used 
in \citetalias{Pittordis_2022} and below, so we downselect later in projected separation.   
The distribution of $a_{inn}$ is uniform in $\log a$ between $1 \au$ to $0.2 a_{out}$, where the outer
limit is an approximate criterion to ensure long-term stability of the system. 
{\newa Binaries closer than $1 \au$ clearly exist, but the velocity-averaging over the anticipated 10-year GAIA 
 baseline greatly suppresses the velocity perturbation - see Sec \ref{sec:closebins} for further details. }  
The orbit eccentricities are chosen independently from the distribution of \citet{Tokovinin_2016},
which is $f(e) = 0.4 + 1.2e$.   Relative orbit inclinations
are random, so a rotation matrix is generated to convert the inner-orbit separations and velocities into the
plane of the outer orbit.  
Masses are drawn from a simplified distribution which is flat in $M$ from $0.01 \msun$ to $0.7 \msun$, then
declining with a $-2.35$ power law above $0.7 \msun$. For each triple we draw three random masses independently
from this distribution, then if $M_3 > M_2$ we swap labels so $M_2 \ge M_3$; 
we then require $M_1, M_2 > 0.4 \msun$ for detectability,  otherwise re-draw a new triplet of
masses. 

After setting up these orbit parameters, we then pick 10 random epochs for each orbit
(i.e. choose mean anomaly uniform in 0 to $2\pi$); solve the Kepler equations to get the true anomaly 
(angle from pericenter) and 
generate a relative 3D velocity and separation from these; we then ``observe" each system at each random time 
from 10 random viewing directions, to produce the relevant observables of 2D sky-projected 
velocities and angular separations. 

We also generate a random distance for each system consistent with the distribution of distances $\le 300 \pc$ 
in PS22, and compute angular separations.   The inner-orbit velocity  $\bv_{inn} \equiv \bv_2 - \bv_3$ is then suppressed 
by a factor $f_{pb}$, the ratio of photocentre-barycenter distance to total separation in the inner orbit. 
There are then two cases  on observations as simulated in PS22; if the inner-orbit angular separation is $\theta_{inn} \le 1 \arcsec$, we assume that GAIA measures ``object 2" as the photocenter of stars 2+3 ; while if $\theta_{inn} \ge 1 \arcsec$
we assume GAIA measures the photocenter as star 2 only, and star 3 is assumed to be resolved but undetected.  Therefore $f_{pb}$ is defined by 
\begin{equation} 
	f_{pb} = \begin{cases} 
		\frac{M_3}{M_2+M_3} - \frac{L_3}{L_2 + L_3}  \qquad  (\theta_{inn} < 1 \text{ arcsec}  ) \\ 
		\frac{M_3}{M_2+M_3}   \qquad  (\theta_{inn} \ge 1 \text{ arcsec} ) 
	\end{cases} 
\end{equation}  
where the $L_{2,3}$ are the model luminosities. 

The observable velocity difference in the wide system is then defined by  
\begin{equation}
	\mathbf{v}_{3D,obs} = \mathbf{v}_{out} - f_{pb} \, \mathbf{R}\mathbf{v}_{inn}
	\label{eq:v3d} 
\end{equation}
where $\mathbf{v}_{out}$ is the outer orbit velocity (star 1 relative to the barycentre of 2+3), and $\mathbf{v}_{inn}$ is
the relative velocity between stars 2+3 in its own plane.  

Then $\bv_{3D,obs}$ is projected to 2-dimensional projected velocity $\Delta v_p$ according to the random viewing direction; the result of this is
a set of 800,000 random triple-system snapshots which include the key observables $\Delta v_{p}$ and $r_p$
along with other orbit parameters.  We then take subsets of these in slices of these observables, and then
test whether or not the presence of star 3 is detectable by a number of methods; these tests are
outlined in the next section. 

\section{Detectability of the third star} 
\label{sec:tripdet} 

Here for each simulated triple-system snapshot, we test whether the third star is detectable by imaging,
astrometry/radial velocity methods, or both or neither. We note that the outer orbits here have separation $\ge 16$ arcsec, 
orbital period $> 10^5$ years and negligible acceleration,  
so we assume our outer ``star 1" has no effect on the detectability; only the parameters
of the inner binary 2+3 are relevant. 
The methods are subdivided into cases of increasing difficulty/observing cost as follows: 

\subsection{Imaging} 
Imaging is optimal for relatively wide inner orbits $r_{p,inn} \ga 30\au$ and when star 3 is above 
the bottom of the main sequence, since it is simple and a positive detection may be obtained in a single epoch.   
For imaging, we test in increasing order of ``observing cost" as follows: first direct seeing-limited imaging, then speckle imaging, then coronography, with detection criteria defined as follows:  
\begin{enumerate}  
	\item  For direct seeing-limited imaging we adopt parameters appropriate for the future Rubin first-year dataset; 
	we adopt a model point-spread function with (pessimistic) 1.5 arcsec seeing and a Moffatt profile with $\beta = -2.5$. 
	We then assume the third star is  detectable if the PSF peak of star 3 exceeds the PSF of star 2 at the centroid
	of star 3 (e.g. assuming 20 percent PSF subtraction accuracy for a 5-sigma detection of star 3). 
	
	We note here that false-positive ``triples" from random background stars are a concern in the case of direct imaging
	at separations $\ga 1\,$arcsec;  however, 
	resolved third stars brighter than $G \la 20$ were already rejected by PS22; for faint stars detectable by Rubin, 
	most should be M-dwarfs or later, and  more distant background stars should mostly have colours / magnitudes 
	inconsistent with an M/L/T dwarf at the measured distance of star 2. 
 {\newa The PS22 sample is limited to galactic latitude$\vert b \vert > 15 \deg$ to 
  minimise background confusion issues, so the probability
   of a random background star at $\le 1 arcsec$ is below 1 percent, except for two small patches 
   near 20 deg from the Galactic center. } 
	
	\item If our star 3 fails to pass the detection threshold for seeing-limited deep imaging above, 
	we next test for speckle imaging; 
	although the contrast is not as good as for AO-assisted coronography (see below), 
	speckle imaging has advantages of low observing overheads (helpful for large samples and short exposures as here), 
	and can detect companions fairly close to the diffraction limit at $I-$band. 
	We assume the quoted performance of the twin instruments Alopeke and Zorro at Gemini N/S respectively 
 ( \citet{Scott_2021}, \citet{Howell_2022} );   
	we approximate the detectable delta-magnitude vs separation as two straight lines as follows: 
	\begin{equation} 
		\Delta m_{max}  = \begin{cases}  0 + 5 (\theta - 0.05)/0.15 \ \text{  if } 0.05 \le \theta \le 0.2 \text{ arcsec} \\ 
			5 + 0.2 (\theta - 0.2)   \ \text{   if } 0.2 < \theta < 1 \text{ arcsec}    
		\end{cases} 
		\label{eq:speckle} 
	\end{equation}  
	Here these 2 lines are slightly pessimistic compared to the actual red-band performance curves in \citet{Howell_2022}. 
	(Note here the outer angular limit of 1 arcsec is set by the small size of the detector window region, due to the need 
	to read-out the window at over 100 frames/sec.) 
	If the angular separation is $0.05 \le \theta \le 1$ arcsec, and the 
	simulated contrast $\Delta m$ of star 3 vs star 2 is below the $\Delta m_{max}$ given by Eq.~\ref{eq:speckle}, 
	we define star 3 as detectable by speckle imaging. 
	Otherwise, we test for coronagraphy. 
	
	\item For coronography, we adopt detection limits similar to the projected performance of the near-future ERIS
	instrument \citep{Davies_2018} on ESO VLT. 
	This is an near-infrared AO-assisted imager/spectrograph including a coronagraph option. 
	Based on Figure~8 of \citet{Davies_2018}, 
	we adopt a contrast limit of $10^{-3}$ to inner working angle of $2.44 \, \lambda/D$, corresponding to $0.14$ arcsec
	at K-band; this contrast limit is rather conservative, as the Figure indicates actual contrast performance 
	closer to $10^{-4}$.  (We note here that dedicated planet-imagers such as SPHERE can achieve much better contrast; however
	these require a very bright primary star for the required AO performance, 
	so most of the PS22 sample are too faint for SPHERE).  
	Assuming an approximate K-band luminosity-mass relation $L_K \propto M^{2.6}$, the $10^{-3}$ contrast 
	translates to a limiting mass ratio $M_3/M_2 \ge 0.07$.
	We also assume a lower mass limit $M_3 > 0.06 \msun$, since old objects below this mass will have
	cooled to the late-T or Y spectral class and be too faint to reliably detect at our median 180 pc distance; most
	L-dwarfs should be detectable if above the contrast limit.  
	In most cases, this lower mass limit is more stringent than the contrast limit.  
\end{enumerate} 

If none of the above tests result in a simulated detection, we define the third star as non-detected in imaging. 

\subsection{Astrometry/RV} 
A substantial fraction of our simulated triples have angular separations too small, or third stars too faint, to be detected in imaging. 
The GAIA astrometry can detect the influence of a third star via the non-linear motion of the photocentre
of stars 2+3, as in e.g. \citet{Belokurov_2020} and \citet{Arenou_2022} for GAIA DR3, i.e. deviations from a 5-parameter fit with position, parallax and constant proper motion.   

Here we assume the 10-year extended mission, since the detectability of uniform acceleration improves with baseline $T$ 
 as $\sim T^{2.5}$ for fixed scanning cadence; so the 10-year extended mission is much superior 
to the baseline 5-year mission for detecting a near-constant acceleration signal.  For the 10-year mission, we assume that the acceleration signal is non-degenerate with the annual parallax signal (this may not be a good approximation if using the 2.75-year baseline of GAIA DR3,  but should be good for the 10-year mission).    

For GAIA astrometric detection, we split according to whether the inner-orbit period is more or less than 10 years.  
If less than 10 years, a full orbit will be seen by GAIA, with a median of 140 visits. 
A relevant parameter here is the single-visit astrometric precision; this varies as a function of G magnitude,
and a fitting formula is given by Eq.~15 of \citet{Kluter_2020} as follows
\begin{eqnarray} 
	\label{eq:sigast} 
	\sigma_{AL} & = & \frac{ 100 + 7.75 \sqrt{ -1.631 + 680.766 \, z+ 32.732 \, z^2} }{ \sqrt{9} } \, \muas 
	\nonumber \\
	z & \equiv & 10^{0.4 \,[max(G, 14) - 15] } 
\end{eqnarray} 
%%  add ref.. arxiv:1911.02584 . 
where $\sigma_{AL}$ is the 1D along-scan astrometric precision for a single focal plane transit, crossing 9 CCDs. 
This gives a value of $76 \muas$ at $G < 14$, rising to $102 \muas$ at $G = 15$ near the median of 
\citetalias{Pittordis_2022}, and $226 \muas$ at $G =17$, 
the faint-end cutoff of the \citetalias{Pittordis_2022} sample.   
For the present case we assume the orbit is reliably detectable if 
\begin{equation} 
	2 \, f_{pb} \, a_{inn} / d \ge 500 \muas 
\end{equation} 
where the left-hand-side is the shift of the photocentre of stars 2+3. 

For inner orbits longer than 10 years, GAIA sees only a partial orbit, and for much longer periods 
this approximates a uniform acceleration.  We compute the angular acceleration $\dot{\mathbf{\mu}}$ 
of the observable photocentre; in this case the difference between the position at mid-mission 
and the mid-point between initial and final positions is  
given by $0.5 \, \dot{\mathbf{\mu}} \, T^2$ for $T = 5\, \text{yr}$; if this exceeds $500 \muas$, 
corresponding to angular acceleration $\dot{\mathbf{\mu}} > 40 \muas \, \yr^{-2}$, we define
the acceleration as significantly detectable. 

Alternatively, a third star may be detected by radial velocity drift of star 2. For these purposes, this is less costly
than planet-detection since we can flag triple systems without requiring full characterisation, and
false-positives are less wasteful, 
so a significant RV shift between just two well-separated epochs would be sufficient to flag a system as triple. This does
still require a minimum of two spectroscopic observations, so is significantly more costly than imaging observations.  

For radial velocity drift, we compute the radial component of acceleration of star 2 due to star 3. We estimate
an observable radial-velocity precision as $0.03 \kms$ per single epoch, which is realistic {\newa for
our typical spectral type FGK stars of moderate to old ages at $G \la 16$ mag, see e.g. \citet{Pasquini_2012}. } (Note that planet-hunting precision can be much better $\sim 0.001 \kms$, but is only achievable for
considerably brighter stars $G \la 12$). 
We then assume that a shift in RV exceeding $0.25 \kms$ over an arbitrary 5-year baseline 
is detectable at high significance, i.e. radial acceleration of star 2 exceeding $0.05 \kms\,{\rm yr}^{-1}$.  

Finally, short-period spectroscopic binaries may be detected via splitting of spectral lines in a single-epoch spectrum, 
or a secondary peak or asymmetry in autocorrelation of a spectrum.  Based on estimates in \citet{Traven_2020}, 
we choose a detection criterion as a radial-velocity difference between stars 2 and 3 
above $20 \kms$, and mass ratio $M_3/M_2 > 0.7$; this only occurs for small and fast inner orbits, and 
turns out to almost never occur in our relevant range of parameter space, so this test is unimportant later. 

We apply all the tests above to each simulated triple snapshot, with results in the next section. 

\subsection{Close inner binaries} 
\label{sec:closebins} 
{\newa 
 In the current study we have not simulated inner binaries closer than $1 \au$. This may be approximately 
 motivated as follows: these are completely unresolved, and over the 10-year baseline of final GAIA, they will complete many  orbits and the photocenter-barycenter motion will average down to a much smaller bias on the barycenter proper motion. 
 If we assume as above that inner orbits with a projected photocenter-barycenter semimajor axis larger than $250 \muas$ are  detectable as astrometric binaries via the GAIA data, only those with photocenter amplitude smaller than this will remain. 
 Taking a simulated example in a pessimistic scenario, with a $\pm 250 \muas$ sinusoidal modulation of period $0.95$ years,  then taking 100 randomly-timed observations over a timespan of 10 years and fitting to uniform linear motion, 
 results in an rms bias on the barycenter proper motion of only $7 \muasy$ or $0.01 \kms$ at 300 pc,  nearly negligible 
 compared to the typical $0.37 \kms$ orbital velocity in a representative wide binary from PS22.  

There is a secondary effect that very close inner binaries, though giving negligible effect on the observed outer
 orbit velocity, do boost the estimated value of $\vtilde$ (see Eq.~\ref{eq:vtilde} below) 
 since they cause an underestimate of the total system mass; 
 the unresolved binary is modelled as a single star based on the combined luminosity, while the true total mass
 is larger, so $\vtilde$ is boosted by a multiplicative factor of $[ M_{sys} (\text{true}) / M_{sys} (\text{est}) ]^{0.5} $.  
 This is most serious for near-equal mass inner binaries, but these should mostly be detectable via their offset 
  above the main sequence in a colour-magnitude diagram; see e.g. \citet{Hartman_2022}. 
    Systems with mass ratio $M_3/M_2 \la 0.7$ produce
 a relatively smaller boost, also the mass-underestimation effect is diluted by the addition of $M_1$ on both numerator 
 and denominator, and by the square-root factor between mass and $\vtilde$.   
 Such systems are in principle detectable by radial velocity measurements since the radial velocity amplitudes will typically
  be rather large; a single-epoch radial velocity measurement will typically deviate from the wide-separation star by many km/sec. 
 
  Overall, the issue of mass underestimation is smaller, and spectroscopic observations are more observationally demanding, 
 compared to the velocity perturbation issues which are the main focus of this paper; we therefore postpone the details to a later study. 
} 

\section{Results and discussion} 
\label{sec:results} 

\subsection{Simulation results} 
For each of the simulated triple snapshots above, we have various observables and detectability results 
defined as above. 
As in \citetalias{Pittordis_2022}, we take slices in outer orbit projected separation $r_p$, with four slices $5.0 - 7.1 \kau$, 
$7.1-10.0 \kau$, $10.0 - 14.1 \kau$, $14.1-20 \kau$.  In \citetalias{Pittordis_2022} the main relative-velocity
observable is the dimensionless parameter 
\begin{equation} 
	\vtilde \equiv \frac{ \Delta v_p }{ v_c(r_p)} \ \ ,  
	\label{eq:vtilde} 
\end{equation} 
where $\Delta v_p$ is the projected velocity difference
of the 2 objects (assuming both stars are at the mean of the two distances from GAIA), and 
$v_c(r_p)$ is the Newtonian circular-orbit
velocity at the observed projected separation, using masses estimated from a main sequence mass-luminosity relation. 
Here we define the simulated $\Delta v_p$ as the 2D projection of $v_{3D,obs}$  from Eq.~\ref{eq:v3d} above, 
i.e. the velocity of star 1 relative to the ``observable centre" of stars 2 and 3.   
The observed histograms of $\vtilde$ were used in the fitting procedure shown in Figures~13 -- 19 of 
\citetalias{Pittordis_2022}.  
We recall that the range $1.0 \le \vtilde \le 1.5$ is particularly important for gravity testing:  the key signature
of MOND is an excessive fraction of pure binaries in this range, and this excess cannot be mimicked by changing the
eccentricity distribution in GR.   

\begin{figure*} 
	\begin{center} 
		\includegraphics[width=16cm]{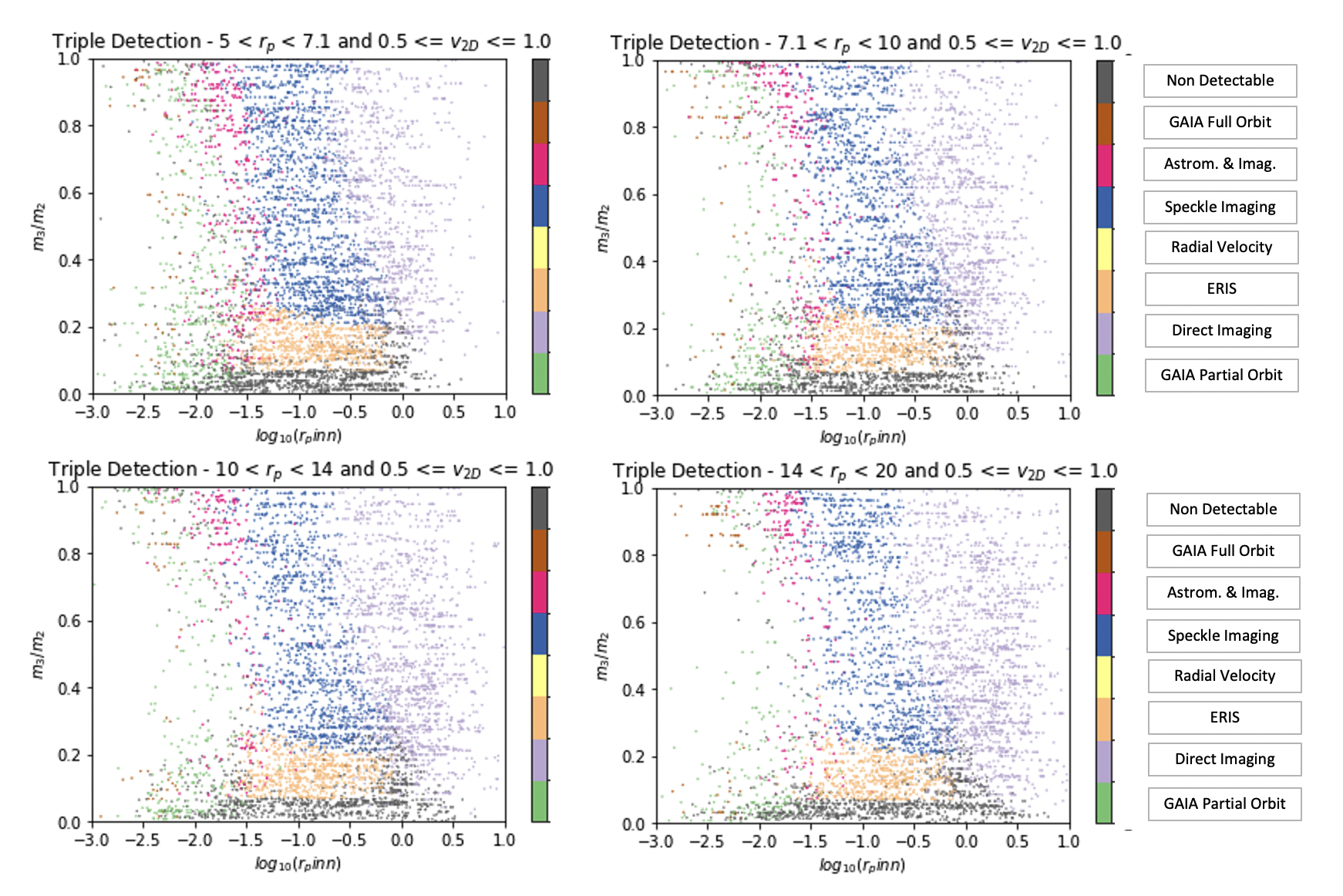} 
		\caption{Scatter plot of mass ratio $M_3/M_2$ (y-axis)
			vs inner-orbit projected separation (log scale, x-axis) for simulated triple systems, 
			in the slice of observable velocity ratio $0.5 < \vtilde < 1.0$.   
			The four panels show four slices in outer-orbit projected separation, as in the legend. 
			Points are colour-coded by detectability of the third object, as labelled in the colour-bar: 
			non-detectable (black), GAIA astrometry (brown for period $< 10 \yr$, green for period $> 10 \yr$); 
			seeing-limited imaging (grey), speckle imaging (blue), ERIS coronagraph (orange). Magenta points
			are detectable by GAIA astrometry and at least one imaging method. } 
		\label{fig:tripdet0510} 
		\includegraphics[width=16cm]{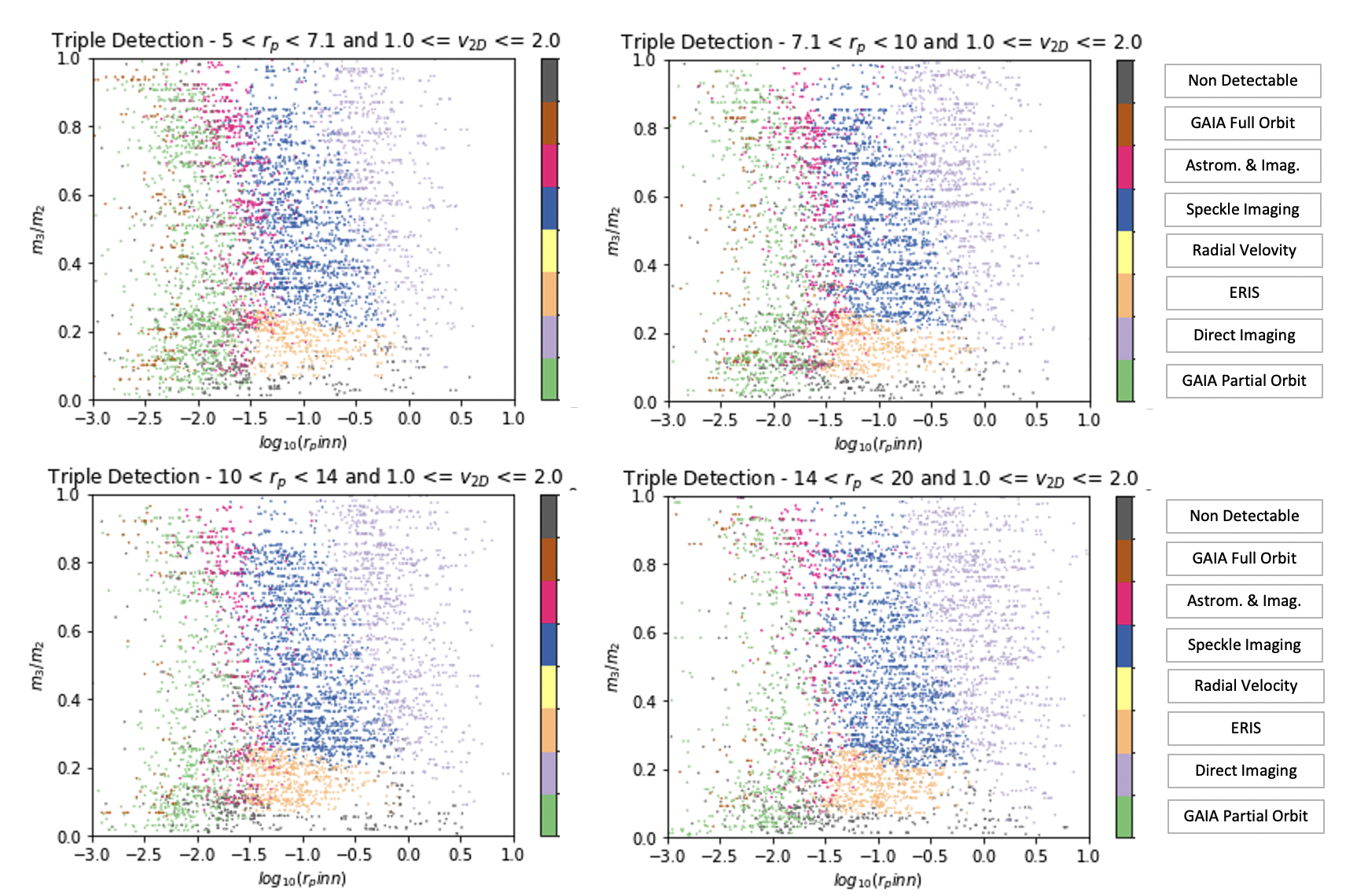} 
		\caption{As Figure~\ref{fig:tripdet0510}, for observable velocity ratio between $1.0 < \vtilde < 2.0$.   }  
		\label{fig:tripdet1020}
	\end{center}
\end{figure*}

\begin{figure*} 
	\begin{center} 
		\includegraphics[width=16cm]{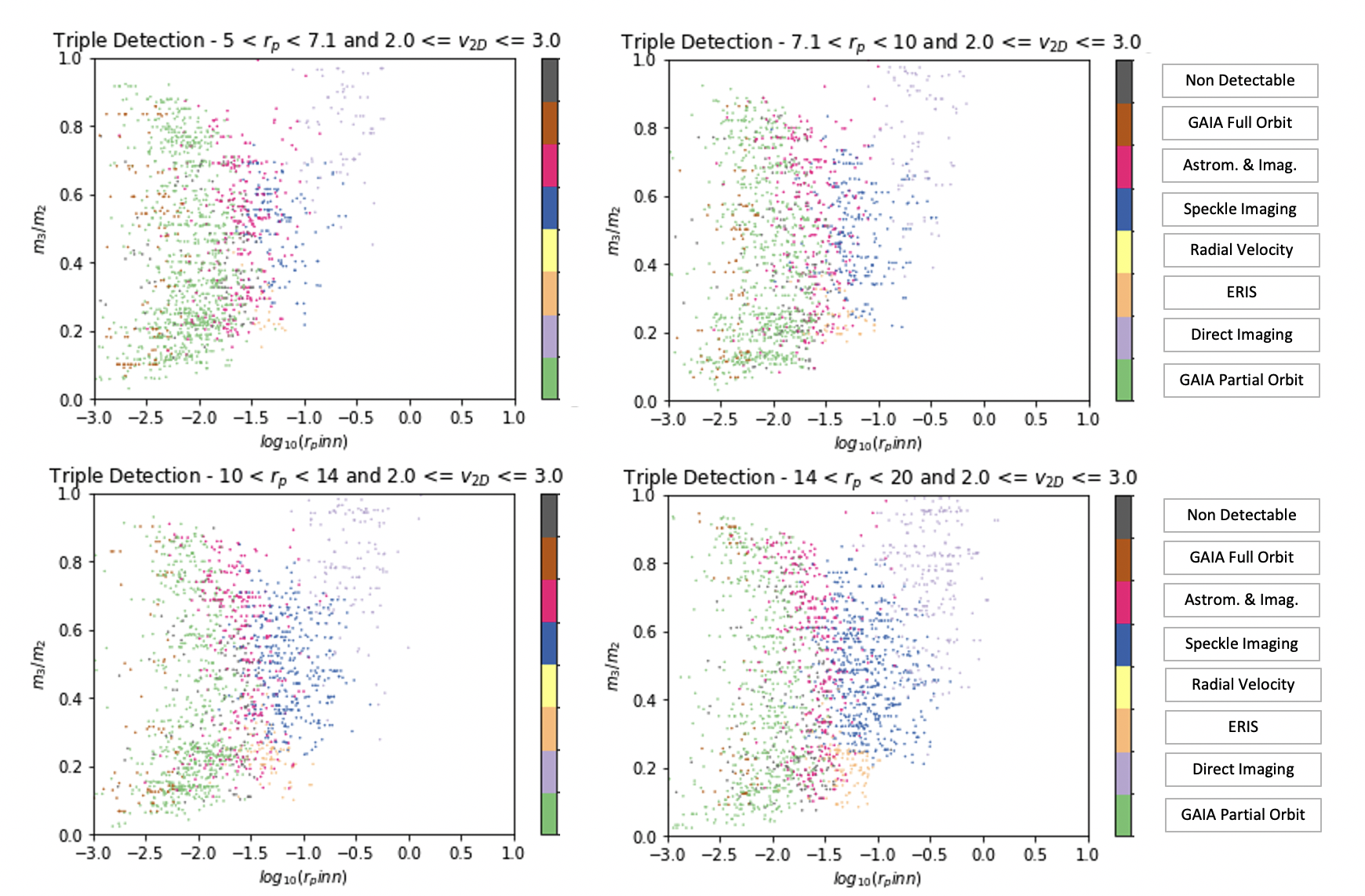} 
		\caption{ As Figure~\ref{fig:tripdet0510}, for observable velocity ratio between $2.0 < \vtilde < 3.0$. } 
		\label{fig:tripdet2030} 
	\end{center}
\end{figure*}

Here, we take slices in $r_p$ and $\vtilde$ as above, 
and then evaluate for each of the simulated systems 
whether the third object is detectable by each of the various methods in Sec.~\ref{sec:tripdet} above. 
Detection percentages by method are shown in Table~\ref{tab:percs}. 
Detection results are shown as scatter plots in Figures~\ref{fig:tripdet0510} - \ref{fig:tripdet2030}, 
relating to the main parameters of the third star, i.e. the mass ratio 
$M_3/M_2$ and the projected separation of the inner orbit $r_{p,inn}$. (Orbit parameters $a_{inn}$ and $e_{inn}$
would be more fundamental, but these will rarely be measurable in practice given the long periods, 
so we focus on direct observables here).  
Plots are colour-coded depending on detectability of the third object by various method(s) as follows: 
black points are undetectable by any of the methods in Section~\ref{sec:tripdet}, 
while magenta points are detectable by {\em both} GAIA astrometric deviations and at least one of the imaging
methods (nearly always speckle imaging or coronography).  Colours other than black or magenta 
are detectable by either some imaging method(s) or astrometry, but {\em not} both. 

\begin{table*} 
		\centering 
	\caption{For simulated triples with outer projected separation $5\kau < r_p < 20\kau$,
		divided in selected $\vtilde$ bins, the table shows the percentage 
		of systems where the third star is detectable by various methods, or none.   
		The subtotal row is the subtotal of rows 1-3, so rows 4-8 sum to almost 100 percent. 
		\label{tab:percs} }  
	\begin{tabular}{ l r r r}  
		& \multicolumn{3}{c}{ Percentage detectable in $\vtilde$ bin: } \\ 
		Method &  $0.5 < \vtilde < 1.0$  & $1.0 < \vtilde < 2.0$  & $2.0 < \vtilde < 3.0 $ \\ 
		\hline 
		GAIA astrometry (full orbit) &  1.0 & 2.1 & 5.0 \\ 
		GAIA astrometry (partial orbit), and not imaging & 6.1  & 16.3 & 43.4 \\ 
		Astrometry/RV and any imaging method & 4.9 & 9.6 & 17.1 \\ 
		\hline 
		Subtotal: Combined astrometry/RV & 12.0 & 28.0 & 65.5 \\   %% subtotal of previous 3 rows 
		Coronagraphic imaging  &  13.7 & 10.8 & 3.8 \\ 
		Speckle imaging  & 26.2 & 30.6 & 18.5  \\ 
		Direct imaging  & 28.7 & 22.4 & 7.8  \\ 
		Not detectable    & 19.4 & 8.0 & 4.4  \\ 
		%% Column total 
		%%% (Number totals:) & 22029 & 19798 & 7030  
		\hline
	\end{tabular} 
\end{table*}

The grey, blue and orange points to the right of these scatter plots show triples detectable by seeing-limited imaging, speckle
imaging and ERIS coronography respectively. As expected given the median distance 180 pc, 
seeing-limited imaging is effective at the largest separations 
$r_{p,inn} \ga 150 \au$, while speckle imaging and coronography are effective down to $\sim 25 \au$.  

To the left of each plot, the brown and green points show inner orbits detectable by GAIA astrometry but none of the
imaging methods; these are coded as period $< 10 \yr$ (brown) or $> 10 \yr$ (green); here longer periods dominate
over shorter, so most astrometric detections will observe a partial orbit.    

As above, the magenta points near the centre show triples detectable by {\em both} GAIA astrometry and 
at least one of the imaging methods (where speckle imaging is the largest contributor). These magenta points are concentrated
near projected separations $r_{p,inn} \sim 30 \au$; this corresponds to angular separations $\sim 0.16 \,\text{arcsec}$ 
at the median distance of \citetalias{Pittordis_2022}, and orbital periods $\sim 150$ years; this implies that GAIA observes around 7 percent of a full orbit and the curvature is sufficient to be detectable, though there is little chance of constraining the overall orbit shape with such a short arc. 

The existence of this magenta overlap region in the centre is a major positive feature of our results: this implies 
that for main-sequence third stars above $M_3 \ge 0.08 \msun$, they are detectable at {\bf any} separation $\ge 1 \au$
by  one or more methods, while very-close inner binaries do not produce appreciable velocity perturbations
 as in Section~\ref{sec:closebins}.   

Here we comment on and explain various features of the scatter plots: 
\begin{enumerate} 
	\item In general the inner-orbit projected separation in each slice 
	has a substantial scatter, caused by the randomness in relative orbit phases, inclinations and viewing angles. 
	\item 
	The divisions between colour regions are relatively clear-cut: this occurs because 90 percent of the sample
	have $90 \pc < d < 300 \pc$ so the scatter in distances is moderate, so projected separations map to
	angular separations with likewise moderate scatter.  
	\item  
	The ``bow-shaped" overall distribution of points is caused by the $f_{pb}$ 
	factor above, which for unresolved inner orbits is maximal at intermediate mass ratios $M_3 /M_2 \sim 0.7$. 
	The velocity perturbation from star 3 depends on the product $f_{pb} \, v_{23}$, so at a given value of $\vtilde$, 
	larger  $f_{pb}$ correlates with slower and wider inner orbits.   
	At wider separations in Figure 3 there is a cloud of points at upper right making a y-shape feature, 
	caused by the larger $f_{pb}$ approaching 0.5 when the third star is assumed separately resolved by GAIA. 
	\item  
	At small separations $\la 10 \au$, nearly all third objects are detectable by GAIA astrometry 
	down to our adopted mass lower limit $0.01 \msun$; 
	even smaller Jupiter-like planets produce reflex motion of star 2 too small to be a concern here. (``Hot Jupiters" can produce
	reflex velocities up to $\sim 0.1 \kms$, but those have periods much less than 1 year, and the resulting perturbation 
	on the mean proper motion of star 2 will average down to a much smaller value over a multi-year GAIA baseline. 
	So Jupiter-mass planets at any separation are not a significant contaminant for the purpose of the gravity test  
	as in \citetalias{Pittordis_2022}. )  
	\item 
	The non-detectable third stars (black points) are primarily cool brown dwarfs below $0.06 \msun$ at projected
	separations $\ga 30 \au$; here the orbital periods are well over 100 years so astrometric accelerations 
	are very small, and also such objects are often too faint for ERIS coronography; L-type brown dwarfs 
	are accessible, but late-T or Y dwarfs are extremely challenging at $\sim 200\pc$ distance.   Such objects in principle may well
	be detectable either with JWST $3 - 5 \mu{\rm m}$ imaging at separations $\sim 0.5 \,\text{arcsec}$,
	or ELT coronagraphy at smaller
	separations; however we have not considered this at the present time, due to the large challenge of getting enough observing
	time on these major facilities for sample sizes of many hundreds of candidate systems; this is deferred for a future study. 
	\item It is notable from Table~\ref{tab:percs} 
	that the fraction of triples detectable with GAIA astrometric acceleration increases steeply with $\vtilde$,
	from 12 percent at $0.5 < \vtilde < 1$ up to 65 percent at $2 < \vtilde < 3$; this may be 
	explained because for fixed angles, orbit phase and masses, orbital velocity scales $\propto r^{-1/2}$ while 
	acceleration scales $\propto r^{-2}$; so inner-orbit acceleration is steeply correlated with velocity. 
	\item 
	It is also notable that the fraction of triples detectable overall (by any method) is quite weakly correlated with outer
	orbit separation $r_p$, but is quite strongly correlated with $\vtilde$; as in Table~\ref{tab:percs}, 
	the overall detection percentage is 80 percent at $0.5 < \vtilde < 1$, rising to 95 percent at 
	$2 < \vtilde < 3$.   This occurs because pure binaries
	must have $\vtilde < \sqrt{2}$ in GR (modulo observational errors), or $\vtilde < 1.65$ in the MOND model chosen in PS22, 
	with over 90\% expected to have values $\vtilde < 1$ in GR.   
	To achieve a value $2 \le \vtilde \le 3$, the inner-orbit perturbation then has to dominate
	the outer-orbit velocity; this means that the lower-right
	portion of Figure~\ref{fig:tripdet2030} is almost empty, since a low-mass $M_3$ at large $r_{p,inn}$
	cannot produce a fast enough perturbation on star 2. 
	The slow slice $0.5 \le \vtilde \le 1$ in Figure~\ref{fig:tripdet0510} includes
	cases where the inner-orbit velocity perturbation is similar or less than the outer-orbit total velocity,  since the simulated (and observed)  $\vtilde$ is a weighted
	resultant of both orbit velocities; so brown-dwarf $M_3$ at large separations commonly end up in this slice (along with 
	a substantial fraction of the pure-binary systems, which were simulated in \citetalias{Pittordis_2022}
	but not repeated here; we simulate pure triples because it is simple to scale results to a general mixture of pure binaries
	and triples, as below). 
\end{enumerate} 

\subsection{Model-dependence of results} 
{\newa 
Our model for the triple population provides a reasonable span of parameter space, but is probably not fully realistic,
 so here we consider the model-dependence.  While the overall statistics of binaries and triples are reasonably well studied 
 (e.g. \citet{Tokovinin_2014}, \citet{Offner_2022} ), the key issue at present is the distribution of inner third stars 
 {\em given} a known wide binary; this is less well constrained due to the limited number of very wide binaries known
  before Gaia.  
 However, the detailed distribution of inner-orbit sizes and eccentricities is probably not too important since
 we see from above that there is good detectability for main sequence third stars at all separations $\ga$ few AU; 
 smaller orbits are either detectable by GAIA astrometry, or are too small to produce significant velocity perturbations
 after time-averaging over 10 years.   

 The key uncertainty is the fraction of brown dwarf companions at $\ga 25 \au$, which dominate our non-detections; 
    our adopted mass model produces 7 percent of third-objects with mass $\le 0.07 \msun$; 
   here third stars above $M_3 > 0.5 \msun$ are under-represented compared to the global average 
   due to the definition $M_3 < M_2$, so a more
 useful comparison value is the ratio of brown dwarfs per M-dwarf in our third-star population, which is 0.13. 
  This may be compared to the  0.176 brown dwarfs per M-dwarf in the RECONS 10 parsec sample \citep{Henry_2018}. 
%% Henry T.J., Jao, Winters etal AJ 155, 265. 

 Ultimately what matters is the fraction of undetected third stars  
  {\em given} an outer wide binary, and this should be clarified by extrapolation of detections from a future observing
 program.   

Ideally in the future we can remove detected triples from the data sample, then 
  model the population of undetected triples and include these in the fitting procedure; details are postponed
  for future work. 

 However, we note that the details may not be critical: in the event that GR is correct and the population of remaining undetected triples is sufficiently small, i.e.  well below the difference between MOND and GR predictions for pure-binary counts, 
 then we may be able to place strong constraints on MOND {\em without} detailed modelling of undetected triples; 
 for example, if the observed  
 population (pure binaries plus undetected triples) in a relevant range around $1.0 \la \vtilde \la 1.5$ 
 turns out to be significantly {\em below} the MOND prediction from pure binaries alone,  then adding a positive number of undetected triples to the MOND prediction can only make the MOND fit worse.   
 Conversely, if the benchmark MOND model of PS22 is actually correct, 
 obtaining acceptable fits assuming GR may demand an unphysically large fraction of undetected triples. 
} 

\subsection{Implications for follow-up observations} 
\label{sec:followup} 
We now turn to the prospects for a realistic observing strategy. One point is that given observed $r_p$ and $\vtilde$ as
in PS22, the potential third-star may scatter over a rather wide region in $M_3, r_{p,inn}$ plane, so there is not much guidance
on which detection method to use; therefore it is best to try all detection methods in ascending order of observing cost; 
but once a system is  proven as a triple by any method, further observations are not strictly necessary.  
To minimise observing cost, it is most economical to defer observations 
until after the GAIA 10-year data release and a year or two of Rubin survey data is available, since these can 
detect a fairly substantial fraction of third-stars with no added observing cost, only moderate software and manpower effort.  
Adding radial-velocity observations appears to add surprisingly little value, since almost all systems showing 
measurable RV drift are  also detectable with GAIA astrometry; even if an RV observing program started now, 
GAIA will pass its 10-year lifespan sooner than our assumed 5-year RV baseline is achieved. 

Also, focusing effort on the faster $\vtilde$ systems is potentially more economical in observing time, since 
they are less numerous in the PS22 sample, the large majority of observed systems in this range are expected to be triples, 
 and can be``discarded" after a detection; while the pure binaries will 
return null results and require observing all the methods to get maximum purity of the survivors. 
Somewhat counter-intuitively, this means that regions of parameter-space containing more triples are
easier for follow-up. 

Combining the four $r_p$ slices above, the PS22 sample (refined with additional RUWE < 1.4 cut) contains $428$ systems with  $5 < r_p < 20 \kau$ and $2 \le \vtilde \le 3$; {\newa in the PS22 model fits, this region of parameter space contains no pure binaries (because the specific chosen instance of MOND produces a maximum $\vtilde \simeq 1.65$)}, but contains approximately 83 percent triples and 17 percent flyby pairs. 
According to the simulated triples above, about 73 percent of the triples in this slice (60 percent of the total) should reveal as such in GAIA or Rubin data.  
This leaves $\sim 180$ systems, $ \sim100$ triples and 80 flybys, to follow-up initially with speckle imaging; this should reveal approximately 60 further triples, leaving only 120 systems or 240 stars to observe with coronographic imaging;  these numbers
look relatively achievable for complete followup.   

A campaign focused on $1 \le \vtilde \le 2$ systems is more directly relevant to the gravity test, since
it is this slice which contains the key difference between the GR and MOND pure-binary distributions 
in the fitting procedure; but this is also more costly in observing time. 
This slice contains 1520 systems in the PS22 sample (with RUWE cut as above) at $5 < r_p < 20 \kau$, and
the PS22 model fit estimates that 40 percent  of these are pure-binary systems; those should obviously return a null result 
for all the third-star tests, and will demand both speckle and coronographic imaging of these to get high
completeness for triples, hence over 1100 stars for a complete sample; this looks possible but a fairly
large-scale observing program.  

For the low-velocity slice $0.5 \le \vtilde \le 1$, this is still more challenging: this slice contains 2523 
systems in the PS22 sample (updated as above), 
and fits predict that 66~percent of these are pure-binary systems; as above, we need to follow-up
all the pure binaries along with the subset of triples which did not reveal a detection in astrometry/direct imaging.  
So this implies $\ga 4000$ stars 
requiring speckle and coronagraphic imaging to get reasonably complete detection of the triples. This is 
very challenging for followup of the entire sample.    

However, note that it is not necessary to detect all the triples individually to obtain a useful gravity test: as long
as we can get a reasonably secure statistical model of the triple population, we can extrapolate this model to 
undetectable third stars, and correct for random-sampling in the fraction of systems actually observed. 
Therefore it seems likely that a weighted campaign is most efficient, observing quasi-random subsamples of systems
with a selection probability varying as a rising function of $\vtilde$ up to some upper limit.  This should help to deliver
an accurate model for the overall population of triples in parameter space. 
Optimising the detailed design of such a program is left to future work. 

For the present, the main positive result from this study is that a high fraction of triple
systems in the PS22 sample of wide-binary candidates 
can be securely detected as triple with a large but realistic observing program.  
This should allow building a robust model for the overall triple population; then a joint fit of binaries, triples and
other systems to a wide-binary sample similar to the PS22 
sample (updated with improved GAIA data) should allow a strong test for MOND-like modified gravity in the medium-term
future.

\section{Conclusions} 
\label{sec:conc} 

We have simulated prospects for observational detection of a third star in the sample of candidate wide-binary systems 
selected and analysed by \citetalias{Pittordis_2022}; this is motivated by the importance of understanding or
removing triple systems to allow a clean test for modified gravity. 

We generated a large sample of simulated hierarchical triple systems with random 
masses, orbit sizes, eccentricities and inclinations, then took ``snapshot" observations of these at random
epochs and viewing angles to produce observables $r_p$, $\vtilde$ as in \citetalias{Pittordis_2022}. 

For each system, we then evaluated
the detectability of the third star using various observing methods including seeing-limited imaging, speckle imaging, 
coronagraphic imaging, GAIA astrometric acceleration, and radial velocity drift, and assessed total detectability 
across parameter space.  
The prospects are encouraging in that a high fraction of triples {\newa where the third star produces a significant velocity perturbation }   
 are detectable as such, increasing with
velocity ratio $\vtilde$ from $\sim 80$ percent at $\vtilde < 1$ to 95 percent at $2 < \vtilde < 3$.  
It is fortunate that the lower separation limit of speckle imaging somewhat overlaps
with the upper limit for GAIA astrometric acceleration, 
so almost all main-sequence third stars producing a velocity pertubation are detectable at any separation; 
non-detections are predominantly cool brown dwarfs at $r_p \ga 25 \au$ which produce too small 
astrometric acceleration and are very faint.  

In principle, a realistic observing program starting soon after the GAIA extended mission 10-year 
data release can strongly constrain the population of triples, and then 
velocity differences in wide binaries should become a strong test for MOND-like theories of modified gravity 
at low accelerations.

%%% The key figs on velocity differences. 

%% force the Figs to come out before the conclusions ... 
% \clearpage

% put this back for final arXiv submission. 
%This is an author-produced, non-copy-edited version of the paper as
%accepted by MNRAS. The version of record is available at 
% DOI:10.1093/mnras/abc9999  . 
%\fi

%%%%%%%%%%%%%%%%%%%%%%%%%%%%%%%%%%%%%%%%%%%%%%%%%%

%%%%%%%%%%%%%%%%%%%% REFERENCES %%%%%%%%%%%%%%%%%% 
%%% RRRRR tag for refs. 

% The best way to enter references is to use BibTeX:

%\bibliographystyle{mnras}
%\bibliography{example} % if your bibtex file is called example.bib

%\bibliographystyle{mnras}
\bibliographystyle{aa_url} % style aa.bst
\bibliography{MSP22_biblio}

	----------------------------------------------------------------

\label{lastpage}
\end{document}